\documentclass[12pt]{article}
\usepackage{amsmath,amssymb,color}
\usepackage{cite,epsf}
\usepackage{pstricks}
\usepackage{color}
\usepackage{hyperref}
\usepackage{tikz}     

\usetikzlibrary{
decorations.pathmorphing 
}

\usepackage{graphicx,array} 
\newcommand{\be}{\begin{equation}}
\newcommand{\ee}{\end{equation}}
\newcommand{\beq}{\begin{equation}}
\newcommand{\eeq}{\end{equation}}
\newcommand{\bea}{\begin{eqnarray}}
\newcommand{\eea}{\end{eqnarray}}

\newcommand{\ba}{\begin{eqnarray}}
\newcommand{\ea}{\end{eqnarray}}

\def\sin{\mbox{sin}}
\def\cos{\mbox{cos}}

\def\log{\mbox{log}}

\begin{document}

\begin{titlepage}
\vspace{10pt}
\hfill
{\large\bf HU-EP-18/01}
\vspace{20mm}
\begin{center}

{\Large\bf  On a new type of divergence for spiky\\[2mm]
 Wilson loops and related entanglement entropies \\[2mm]

}

\vspace{45pt}

{\large Harald Dorn 
{\footnote{dorn@physik.hu-berlin.de
 }}}
\\[15mm]
{\it\ Institut f\"ur Physik und IRIS Adlershof, 
Humboldt-Universit\"at zu Berlin,}\\
{\it Zum Gro{\ss}en Windkanal 6, D-12489 Berlin, Germany}\\[4mm]

\vspace{20pt}

\end{center}
\vspace{10pt}
\vspace{40pt}

\centerline{{\bf{Abstract}}}
\vspace*{5mm}
\noindent
We study the divergences of  Wilson loops for a contour with a cusp of zero opening angle, combined with a nonzero discontinuity of its curvature.
The analysis is performed in lowest order, both for weak and strong coupling. Such a spike contributes a leading divergent term
proportional to the inverse of the square root of the cutoff times the jump of the curvature. As nextleading term appears a logarithmic one
in the supersymmetric case, but it is absent in QCD. The strong coupling result, obtained from minimal surfaces in AdS via holography, can be used 
also for applications to entanglement entropy in (2+1)-dimensional CFT's.

\vspace*{4mm}
\noindent

\vspace*{5mm}
\noindent
   
\end{titlepage}
\newpage


\section{Introduction}
Wilson loops for smooth contours in non-supersymmetric gauge theories, besides a linear divergence proportional to the length,
do not require any further renormalisation beyond that of the coupling constant \cite{Polyakov:1980ca, Dotsenko:1979wb}. For the local supersymmetric generalisation in ${\cal N}=4$ super Yang-Mills \cite{Rey:1998ik, Maldacena:1998im }  the situation is even more comfortable: the coupling is not renormalised, and the linear divergence cancels between the gauge boson and scalar contribution \cite{Drukker:1999zq}.

For contours with cusps or self-intersections each of these singular points generates renormalisation Z-factors \cite{Polyakov:1980ca,Brandt:1981kf}. The corresponding cusp anomalous dimension depends on the cusp angle 
\footnote{With the convention: opening angle is $\pi-\vartheta$.} $\vartheta$ and the coupling constant
and has been calculated in the $80$-ies up to perturbative two loop level \cite{Polyakov:1980ca, Knauss:1984rx, Korchemsky:1987wg}. Later on it turned out to be related to various other physical situations, and with the advent of AdS-CFT holography
 it became one of the most studied quantities from both the weak as well as the strong coupling side. It is now available up
to three loops both for ${\cal N}=4$  SYM and QCD  \cite{Correa:2012nk, Grozin:2014hna}. For strong coupling one has the
leading and next leading contribution \cite{ Drukker:1999zq, Forini:2017whz}. Of special interest is also the limit
of large imaginary angle \cite{Korchemsky:1987wg}. It plays a crucial role for scattering amplitudes and the dimensions of large spin operators,  and using techniques of integrability even an interpolation between weak and strong coupling has been found \cite{Eden:2006rx, Beisert:2006ez}.

The minimal string surfaces, one has to consider for the strong coupling evaluation, play still another prominent role. They
carry all the information needed for the strong coupling evaluation of entanglement entropies via holography in
(2+1)-dimensional conformal field theories \cite{Ryu:2006ef, Ryu:2006bv}. The cusp anomalous dimension appears then
as the coefficient of the logarithmic divergence due to a cusp in the boundary of a subregion in 2-dimensional static  space.
 In the following we mainly use the Wilson loop language, translations into the context of entanglement entropy should be straightforward.

Besides the already mentioned limit for infinite imaginary angle, the limits $\vartheta\rightarrow 0$ and $\vartheta\rightarrow\pi$
have been studied in detail. The first limit corresponds to the approach to a smooth contour, and therefore the cusp anomalous dimension vanishes. Its approach to zero is of quadratic order in $\vartheta$,  with a coefficient related to the Bremsstrahlung
of a heave charge  \cite{Correa:2012at}. In the other limit the cusp anomalous dimension diverges proportional to $1/(\pi-\vartheta)$ with a coefficient, which can be identified with the static quark-antiquark potential on the sphere \cite{Correa:2012nk}, for a related statement see also \cite{Drukker:2011za}.

If one looks at the lowest order Feynman diagrams responsible for the  cusp logarithmic  divergence, one realises that this divergence
is absent if one puts $\vartheta=\pi$ before the removal of the regularisation, i.e. the limit $\vartheta\rightarrow\pi$ does
not commute with renormalisation \cite{Dorn:1986dt}.

$\vartheta =\pi$ corresponds to a jump of the tangent vector of the contour
to its additive inverse. One situation, where this is relevant, appears for
a contour which runs on a piece forward and backward and is connected to the
issue of zigzag symmetry \cite{Polyakov:1997tj, Drukker:1999zq}.

Our main interest in this paper concerns the case, for which one has $\vartheta =\pi$
{\it and} a finite nonzero jump in the curvature of the contour, i.e. a case where
the cusp degenerates to a spike.\footnote{Our interest in this situation has been
triggered by a recent paper which studies a different type of spikes, those built by spirals \cite{Pastras:2017fsy}.}$~$\footnote{To fix nomenclature, following parts of the physical literature, we use the word cusp for a corner with nonzero opening angle $(\pi -\vartheta)$ and spike for a cusp with zero opening angle. } We will find a new type of divergence which depends on the jump
in curvature. In this respect the $\vartheta=\pi$ case is very special, because
for $0\leq\vartheta<\pi$ any dependence of the divergence on other local
quantities beyond $\vartheta$ has been excluded, both in small coupling perturbation
theory \cite{Dorn:1986dt} as well as in the holographic treatment for strong coupling \cite{Dorn:2015bfa}.

In the following section 2 we consider the lowest order for weak coupling, both
for QCD and  ${\cal N}=4$ SYM. Section 3 is devoted to a study of minimal
surfaces in AdS, relevant for the strong coupling behaviour in  ${\cal N}=4$ SYM.
After the concluding section 4 the paper is completed by some technical appendices.  
\section{Lowest order perturbation theory}
The Euclidean local supersymmetric Wilson loop for a closed contour parameterised by
$x^{\mu}(\sigma)$ is given by \cite{Rey:1998ik, Maldacena:1998im, Drukker:1999zq}
\beq
W~=~\frac{1}{N}\big \langle \mbox{tr}~ P\mbox{exp} \int \big (iA_{\mu}\dot x^{\mu}+\vert \dot x\vert \phi_I\theta^I\big )d\sigma \big \rangle~. \label{w-loop}
\eeq
The coupling to the scalars $\phi_I$ is controlled by a contour $\theta^I(\sigma)$ 
on $S^5$. Here we restrict ourselves to constant $\theta^I$. The extension to
nontrivial contours and cusps or spikes on $S^5$ would be straightforward. In the 
non-supersymmetric case the scalars are absent. 
The divergence of interest for QCD is given by the two diagrams in figure 1. In the supersymmetric case the analog diagrams
with scalar propagator have to be added.


\vspace*{2cm}

\begin{tikzpicture}[scale=0.5]
\draw [->,very thick] (0mm,0mm) arc (270:380:50mm);
\draw [-<,very thick] (0mm,0mm) arc (270:340:100mm);
\draw [very thick,red ,decorate,decoration={snake,amplitude=.5mm,segment length=1.5mm,post length=.2mm}] (57.3576mm,18.0848mm)--(45.3154mm,28.8691mm);
\draw [->,very thick] (150mm,0mm) arc (270:380:50mm);
\draw [-<,very thick] (150mm,0mm) arc (270:340:100mm);
\draw [very thick,red ,decorate,decoration={snake,amplitude=.5mm,segment length=1.5mm,post length=.2mm}] (178.6788mm,9.0424mm)--(199.2405mm,58.6824mm);
\end{tikzpicture}\\[10mm]
Figure 1: {\it The lowest order gluon diagrams contributing to the spike divergence. The\\
$~~~~~~~~~~~~$contour near the spike is shown in black with arrows pointing to increasing\\
 $~~~~~~~~~~~~$contour parameter. Red wavy lines are gluon propagators.}\\[2mm]

At this point a comment is in order, explaining why the left diagram does not
give the complete answer and why then the right diagram has to be included, but not
the analogous diagram for the other leg of the spike.

The diagrams with both ends of the propagator on one and the same
smooth piece of the contour generate a logarithmic end point contribution. It has to be attributed to the endpoint
belonging to the smaller value of the contour parameter $\sigma$.\footnote{In QCD these end point contributions are responsible for an anomalous dimension of Wilson operators for {\it open}
contours \cite{Gervais:1979fv,Arefeva:1980zd,Craigie:1980qs,Aoyama:1981ev}.}
 For the right diagram in figure 1 this point is the tip of
our spike under consideration, for the analogue on the other leg it would be the foregoing cusp or spike.

 For the case of standard cusps this procedure guarantees
the vanishing of the cusp anomalous dimension in the smooth limit.
A convenient
setting for discussing all the renormalisation issues of the non-local Wilson operators in the language of local objects
is provided by the use of an one dimensional  auxiliary field living on the contour, for a review see \cite{Dorn:1986dt} and references therein.

The starting point for the analysis of the contribution of the spike to the 
divergences is then ($g$ coupling, $C_F$ quadratic Casimir for the fundamental representation of $SU(N)$ and $+\dots$ for order $g^2$ contributions from outside a vicinity of the spike)
\beq
\mbox{log}~ W~=~\frac{g^2C_F}{4\pi^2}\Big (I_{\mbox{\scriptsize sc}}^{\mbox{\scriptsize sp}}~+~I_{\mbox{\scriptsize sc}}^{\mbox{\scriptsize end}}-~I_{\mbox{\scriptsize gl}}^{\mbox{\scriptsize sp}}~-~I_{\mbox{\scriptsize gl}}^{\mbox{\scriptsize end}} ~+~\dots ~\Big )~+~{\cal O}(g^4)~,\label{wloop2}
\eeq
with
\bea
I^{\mbox{\scriptsize sp}}_{\mbox{\scriptsize sc}}&=&\int ^{\sigma_s}d\sigma_1\int_{\sigma_s}d\sigma_2~\frac{1}{D_{1,2}}~,~~~~~I^{\mbox{\scriptsize end}}_{\mbox{\scriptsize sc}}~=~\int _{\sigma_s}d\sigma_2\int_{\sigma_s}^{\sigma_2}d\sigma_1~\frac{1}{D_{1,2}}~,\\[2mm]
I^{\mbox{\scriptsize sp}}_{\mbox{\scriptsize gl}}&=&\int ^{\sigma_s}d\sigma_1\int_{\sigma_s}d\sigma_2~\frac{\dot{\vec x}_1\cdot\dot{\vec x}_2}{D_{1,2}}~,~~~~~I^{\mbox{\scriptsize end}}_{\mbox{\scriptsize gl}}~=~\int _{\sigma_s}d\sigma_2\int_{\sigma_s}^{\sigma_2}d\sigma_1~\frac{\dot{\vec x}_1\cdot\dot{\vec x}_2}{D_{1,2}}~,\label{sp}\\[2mm]
D_{1,2}&=&(\vec x(\sigma_1)-\vec x(\sigma_2))^2~+~a^2~.
\eea
For UV regularisation we  have introduced the parameter $a$. As contour parameter
has been chosen the geometrical length, i.e. $\vert \dot x\vert =1$, $\sigma_s$
is its value at the tip of the spike. 

The expansion of $\vec x(\sigma_1)$ and $\vec x(\sigma_2)$ near the tip of the spike
for the left diagram in fig.1 is
\bea
\vec x(\sigma_1)&=&\vec x(\sigma _s)~+~\vec t~ (\sigma_1 -\sigma_s)~+~\vec k_1\frac{(\sigma_1 -\sigma_s)^2}{2}~+~\dots~,~~~~~\sigma_1\leq\sigma_s~,\nonumber\\
\vec x(\sigma_2)&=&\vec x(\sigma _s)~-~\vec t~ (\sigma_2 -\sigma_s)~+~\vec k_2\frac{(\sigma_1 -\sigma_s)^2}{2}~+~\dots~,~~~~~\sigma_2\geq\sigma_s~.
\eea
The appearance of $\vec t$ and $-\vec t$ in the first and second line is due to our choice $\vartheta =\pi$ for the cusp angle. For the right diagram the expansion looks similar, but instead of $-\vec t$ in the second line one has $+\vec t$ and both $\sigma_1$ and $\sigma_2$ are larger then $\sigma_s$.

This implies for the nominators in \eqref{sp}: $\mp 1 + {\cal O}(\sigma_1-\sigma_2)$ for the first and second integral, respectively. 
The order ${\cal O}(\sigma_1-\sigma_2)$ does not contribute
to divergences for $a\rightarrow 0$, hence
\beq
I^{\mbox{\scriptsize sp}}_{\mbox{\scriptsize gl}}~=~-I^{\mbox{\scriptsize sp}}_{\mbox{\scriptsize sc}}~+~{\cal O}(a^0)~,~~~~~~I^{\mbox{\scriptsize end}}_{\mbox{\scriptsize gl}}~=~I^{\mbox{\scriptsize end}}_{\mbox{\scriptsize sc}}~+~{\cal O}(a^0)~.\label{7}
\eeq

We now discuss the calculation of the divergent part of $I^{\mbox{\scriptsize sp}}_{\mbox{\scriptsize sc}}$, the end point integral can be taken from the literature.
After changing variables $\sigma_1-\sigma_s=-\tau_1~,~~\sigma_2-\sigma_s=\tau_2$
and introducing polar coordinates in the $(\tau_1,\tau_2)$-plane we get
\beq
I^{\mbox{\scriptsize sp}}_{\mbox{\scriptsize sc}}~=~\int_0^l\frac{dr}{r}\int _0^{\pi/2}~d\varphi~\frac{1}{1-\sin 2\varphi +\frac{r^2}{4}(\vec k_1~\cos ^2\varphi -\vec k_2~\sin^2\varphi)^2 +a^2/r^2}~+~~{\cal O}(a^0)~.\label{spsc}
\eeq
Here use has been made of $\vec t~\vec k_j=0$ for $j=1,2$. The ${\cal O}(a^0)$ takes
notice of finite terms arising from higher terms in the expansion
of $\vec x(\sigma)$. $l$ is an auxiliary parameter defining a certain vicinity
of the spike, it has no effect on the divergent terms.

The further evaluation of this integral is performed in appendix A, and from there
we get
\beq
I^{\mbox{\scriptsize sp}}_{\mbox{\scriptsize sc}}~=~\frac{\sqrt{\pi}~(\Gamma (1/4))^2}{2~\sqrt{2a~\vert \vec k_{1}-\vec k_2\vert }}~+~\log~ a~+~{\cal O}(a^0)~.\label{9}
\eeq
For the total  spiky contribution to the divergence, we have to combine this with the known
\beq
I^{\mbox{\scriptsize end}}_{\mbox{\scriptsize sc}}~=~\frac{\pi l}{2a}~+~\mbox{log}~a~+~{\cal O}(a^0)~.\label{10}
\eeq
The $1/a$-term is the contribution to the well-known overall linear divergence proportional to the length of the contour. It has nothing to do with the effect of the
spike. Then, putting together \eqref{wloop2},\eqref{7},\eqref{9},\eqref{10}, we 
get for the contribution of the spike
to the Wilson loop in QCD 
\beq
\mbox{log}~W_{\mbox{\scriptsize QCD}}\Big \vert_{spiky}~=~\frac{g^2C_F}{4\pi^2}~\left (\frac{\sqrt{\pi}~(\Gamma (1/4))^2}{2~\sqrt{2a~\vert \vec k_{1}-\vec k_2\vert }}~+~{\cal O}(a^0)\right  )~+~~{\cal O}(g^4)~.\label{weak-QCD} 
\eeq
For the supersymmetric case and smooth path on $S^5$ one finds instead
\beq
\mbox{log}~W_{\mbox{\scriptsize SYM}}\Big \vert_{spiky}~=~\frac{g^2C_F}{4\pi^2}~\left ( \frac{\sqrt{\pi}~(\Gamma (1/4))^2}{\sqrt{2a~\vert \vec k_{1}-\vec k_2\vert }}~+~2~\log~ a~+~{\cal O}(a^0)\right )~+~~{\cal O}(g^4)~.\label{weak-SYM}
\eeq
The absence of a logarithmic term in QCD and its presence in SYM we interpret
as somehow related to the presence or absence of zig-zag symmetry. 

In dimensional regularisation the logarithmic term corresponds as usual  to a pole at dimension 4. The $1/\sqrt a$ term
corresponds to a pole at dimension 3.5.

\section{Holographic evaluation for strong coupling}
For large $N$ and strong 't Hooft coupling $\lambda=g^2N$ in ${\cal N}=4$ SYM one has the holographic formula \cite{Maldacena:1998im}
\beq
\mbox{log}~W~=~-\frac{\sqrt{\lambda}}{2\pi}~A~,\label{loop-malda}
\eeq
with $A$ denoting the area of the minimal surface in AdS approaching the Wilson loop contour on the boundary. 

For the application to entropies in $(2+1)$-dimensional CFT's  this area  gives at strong coupling
up to a factor the entanglement of the two-dimensional region enclosed by the contour \cite{Ryu:2006bv,Ryu:2006ef}.

We do not attempt to solve the difficult task of finding the minimal surface for the generic spike situation, discussed
in the previous section. Instead we generate a special spiky situation by conformal transformation of a well-known
explicit solution, that for the case of two parallel straight lines \cite{Maldacena:1998im}.\footnote{Conformal transformations of this case, combined with T-dualities have been studied recently in \cite{dekel}.} Putting these two lines away from the
origin of our coordinates, after inversion on the origin we get two circles of different radius, touching each other  at one point.
This procedure yields a contour with two spikes with common tips. The small subtleties for the comparison with the 
results in section 2,  due to this touching of the two tips,  will be discussed in appendix C.
The situation is illustrated in Figure 2.

Let us now consider the  two parallel straight lines of distance $L$, located parallel to the $x_1$ axis and crossing the $x_2$ axis
at $x_2= M\pm L/2,~~(M-L/2>0)$. In Poincar$\acute{ \mbox{e}}$ coordinates ($x_1,x_2,z,$ with $z=0 $ as boundary, $ds^2=(dz^2+dx_1^2+dx_2^2)/z^2$) the related minimal surface  is given by  \cite{Maldacena:1998im}
\bea
z(\sigma,\tau)&=& r(\sigma)~,~~~~~~~~~~~~~~~~~~\sigma\in(-L/2,L/2)~,~~\tau\in(-\infty,\infty)~,\nonumber\\
x_1(\sigma,\tau)&=&\tau~,~~~~x_2(\sigma,\tau)~=~M~+~\sigma~.\label{orig-surf}
\eea 
The function $r(\sigma)$ is defined via
\beq
r(-\sigma)~=~r(\sigma)~~~\mbox{and}~~~~\sigma~=~r_0~\int _{\frac{r(\sigma)}{r_0}}^1\frac{y^2dy}{\sqrt{1-y^4}}~~~\mbox{for}~0<\sigma<\frac{L}{2}~,\label{sigma-r}
\eeq
with
\beq
L~=~2r_0\int_0^1\frac{y^2dy}{\sqrt{1-y^4}}~=~\frac{(2\pi)^{3/2}~r_0}{\big (\Gamma(\frac{1}{4})\big)^2}~. \label{r0L}
\eeq
Now we apply the AdS isometry
\beq
x_{\mu}~\mapsto~\frac{x_{\mu}}{x^2+z^2}~,~~~~~~z~\mapsto ~\frac{z}{x^2+z^2}~, \label{trafo}
\eeq
which on the boundary just yields the conformal inversion on the origin. Then the image of \eqref{orig-surf} under \eqref{trafo} is
\bea
x_1&=&\frac{\tau}{\tau^2+(M+\sigma)^2+r^2(\sigma)}~,\nonumber\\
x_2&=&\frac{M+\sigma}{\tau^2+(M+\sigma)^2+r^2(\sigma)}~,\nonumber\\
z&=&\frac{r(\sigma)}{\tau^2+(M+\sigma)^2+r^2(\sigma)}~.\label{image}
\eea
\begin{figure}[h!]
 \centering
\includegraphics[width=14cm]{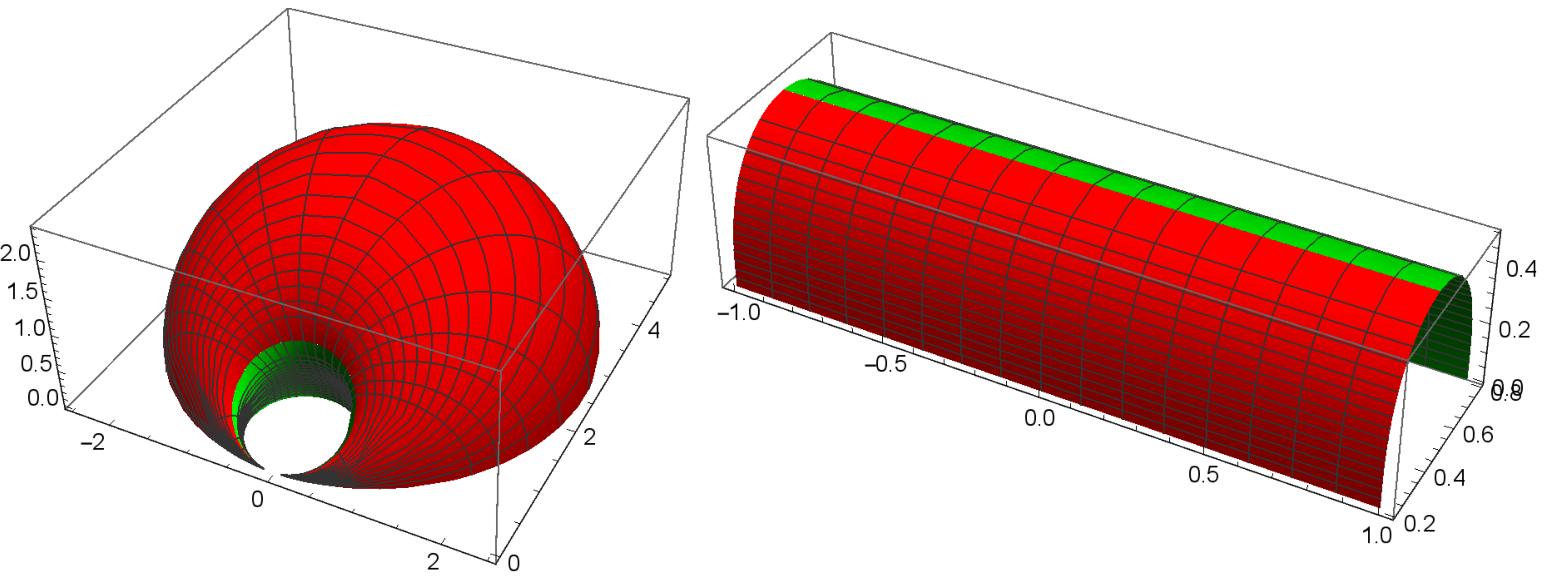}
\\[5mm]
Figure 2: {\it $r_0=0.5, ~~M=0.5$, note the different scales. The green and red pieces
correspond to positive and negative values of $\sigma$, respectively.
}  
\label{fig2}
\end{figure}
\newpage
\noindent
   Figure 2 shows both the original (right) and the image (left) for a special choice of $M$ and $r_0$.

The wanted regularised area $A_{\epsilon}$ is then given by cutting the surface \eqref{image} at $z=\epsilon$ and taking into account
only the piece $z>\epsilon$. Since \eqref{trafo} inside AdS is an isometry, we can calculate this area also on the original
surface \eqref{orig-surf} whose induced metric is simpler. 

The boundary of the region on \eqref{orig-surf}, which has to be taken into account, is then given by
\beq
\epsilon~=~\frac{r(\sigma)}{\tau^2+(M+\sigma)^2+r^2(\sigma)}~.\label{cut}
\eeq
Examples of the corresponding curves on \eqref{orig-surf} are shown in figure 3 for three different values of $\epsilon$.
\begin{figure}[h!] 
 \centering
 \includegraphics[width=12cm]{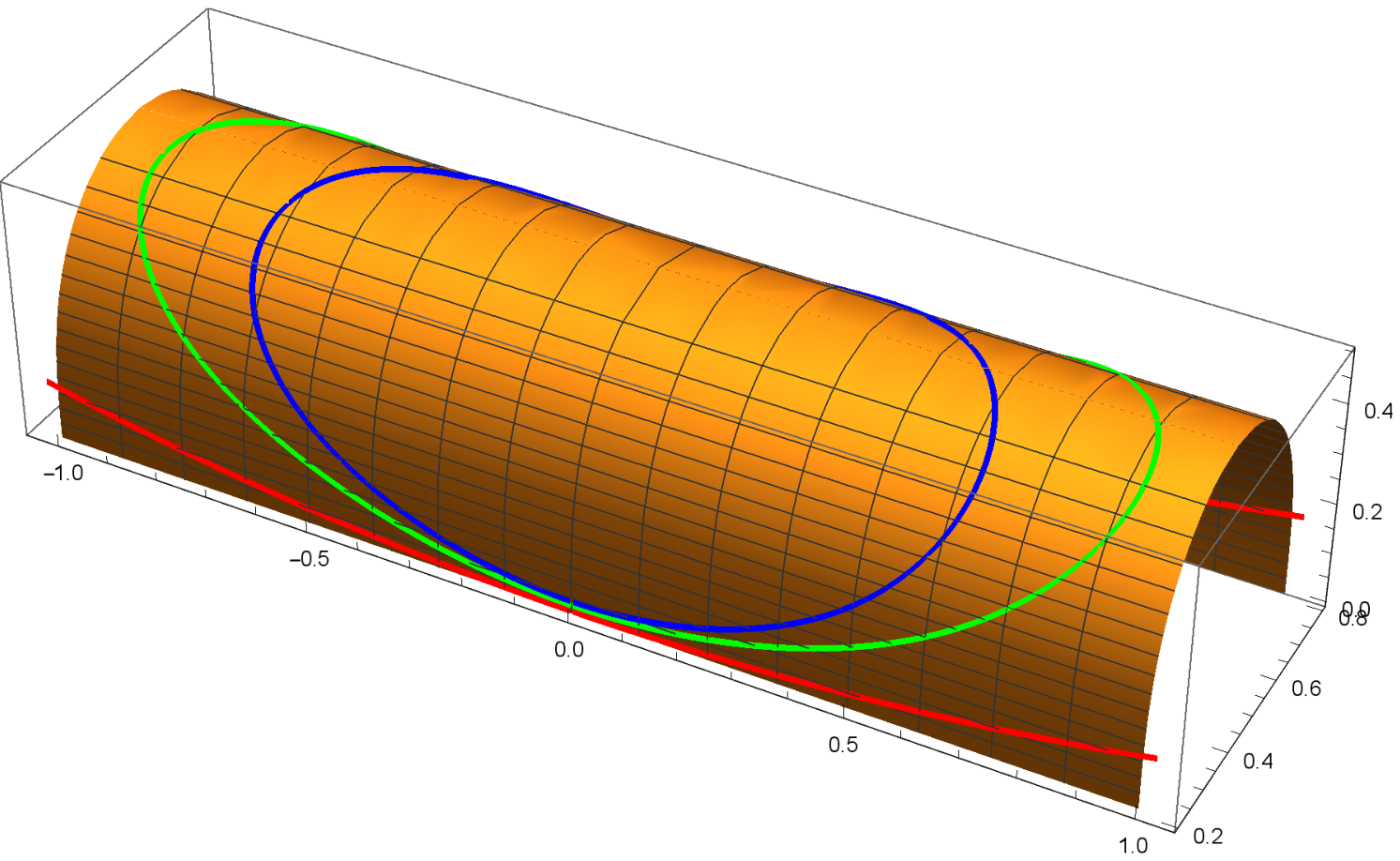} 
\\[5mm]
Figure 3: {\it $r_0=0.5, ~~M=0.5$, red, green, blue: $\epsilon =0.1,~0.4,~0.6$} 
\label{fig3}
\end{figure}
The integration for $A_{\epsilon}$ is performed over all $\sigma,~\tau$ for which the r.h.s. of \eqref{cut} is larger than $\epsilon$.
The integrand, due to the induced metric on \eqref{orig-surf},  is given by $\sqrt{1+r'^2}/r^2$.  This is independent of $\tau$, and the
corresponding conservation law allows to simplify to  $r_0^2/r^4$.  This  yields
\beq
A_{\epsilon}~=~r_0^2~\int \frac{1}{r^4(\sigma)}~d\sigma d\tau~.
\eeq
For the further evaluation it is convenient to change the integration variable $\sigma$ to $r$. Since the relation between $r$ and $\sigma$ in \eqref{sigma-r}
is one to two, one has to split the integral in pieces with $\sigma>0$ and $\sigma<0$. Then with $\sigma(r)\geq 0$ defined by the integral in \eqref{sigma-r}
we get 
\beq
\frac{d\sigma}{dr}~=~-\frac{r^2}{r_0^2}~\frac{1}{\sqrt{1-(\frac{r}{r_0})^4}}~,
\eeq
\beq
A_{\epsilon}~=~A_{\epsilon}^+~+~A_{\epsilon}^-~,\label{A-Apm}
\eeq
\beq
A_{\epsilon}^{\pm}~=~\int _{{\cal B}^{\pm}_{\epsilon}}~\frac{1}{r^2\sqrt{1-(\frac{r}{r_0})^4}}~dr d\tau~.
\eeq
The integration regions ${\cal B}^{\pm}_{\epsilon}$ are defined by
\beq
\frac{r}{\tau ^2+(M\pm \sigma(r))^2+r^2}~>~\epsilon~.
\eeq
For each $r\in {\cal B}^{\pm}_{\epsilon}$ the $\tau$-integration runs between
$\pm\sqrt{r/\epsilon -(M\pm\sigma(r))^2-r^2}$. Since the integrand is independent of
$\tau$ we get
\beq
A_{\epsilon}^{\pm}~=~\frac{2~r_0^2}{\sqrt{\epsilon}}~\int_{r_{\epsilon}^{\pm}}^{r_0}\frac{\sqrt{r-\epsilon~ (M\pm \sigma(r))^2-\epsilon~ r^2}}{r^2~\sqrt{r_0^4-r^4}}~dr~. 
\label{Aeps}\eeq
The lower boundaries $r_{\epsilon}^{\pm}$ are defined as solutions of
\beq
r_{\epsilon}^{\pm}~-~\epsilon~\big ((M\pm \sigma(r_{\epsilon}^{\pm}))^2+(r_{\epsilon}^{\pm})^2  \big )~=~0~.\label{region}
\eeq
These integrals and their behaviour for $\epsilon\rightarrow 0$ are discussed in some detail in appendix B.
 It results in \eqref{Afinal}, i.e.
\beq
A_{\epsilon}^{\pm}~=~\frac{\pi}{\epsilon(M\pm L/2)}~-~\frac{8\sqrt{2}~\pi^{7/4}\sqrt{2\sqrt{2}+3}}{(\Gamma(\frac{1}{8}))^2}~\frac{1}{\sqrt{\epsilon L}}~+~{\cal O}(\epsilon ^0)~.\label{Afinalmain}
\eeq
To compare with the results in section 2, we have to relate  $L$ and $M$  to the length of our contour and to the jump of the curvature at the spikes.
 The radii of the two circles are
\beq
R_{1\vert 2}~=~\frac{1}{2(M\pm\frac{L}{2})}~.
\eeq
This implies for the circumferences $l_1,~l_2$ and the jump in curvature
\beq
l_{1\vert 2}~=~\frac{\pi}{M\pm\frac{L}{2}}~,~~~~~~~~\vert\vec k_1-\vec k_2\vert~=~2L~.
\eeq
Then with \eqref{Afinalmain}, \eqref{A-Apm} we get
\beq
A_{\epsilon}~=~\frac{l_1+l_2}{\epsilon}~-~\frac{32~\pi^{7/4}\sqrt{2\sqrt{2}+3}}{(\Gamma(\frac{1}{8}))^2}~\frac{1}{\sqrt{\epsilon \vert\vec k_1-\vec k_2\vert}}~+~{\cal O}(\epsilon ^0)~.\label{spike-root}
\eeq
The usual $1/\epsilon$ divergence proportional to the length of the contour is not
influenced by the presence of the spikes. Since we have two spikes, the $1/\sqrt{\epsilon}$ divergence to be attributed to {\it one} spike is half of the corresponding term in 
\eqref{spike-root}.

The holographic calculation is relevant for the supersymmetric case. Therefore, it is 
at first sight amazing that the divergence has no logarithmic piece, while  
for weak coupling there is one \eqref{weak-SYM} (but not for QCD \eqref{weak-QCD}).

In appendix C we show, that the weak coupling calculation for the contour of this
section, with its two touching spikes, yields both for QCD and SYM no 
logarithmic term.

What concerns the divergence related to a single spike it remains open, whether
a logarithmic term present at weak coupling survives at strong coupling or
has a coefficient approaching zero. If the absence of a logarithmic term
is related to zigzag symmetry, as argued in the introduction, the restoration
of this symmetry for strong coupling, advocated in \cite{Drukker:1999zq}, would favour the latter.
\section{Conclusions}
We have analysed the divergent contribution to Wilson loops (entanglement entropies)
due to a class of spikes of the contour, those with cusp
angle equal to $\pi$ and with a nonzero jump of the curvature. For weak coupling
the analysis has been done both for QCD and SYM. At strong coupling for
${\cal N}=4$ SYM. 
Both for weak and strong coupling we found in lowest order a divergence 
proportional to
the inverse square root of the product of the dimensionful short distance cutoff and the
discontinuity of the curvature. 

Let us identify  the cutoffs for the weak and
strong coupling case, i.e. $a=\epsilon$ and assume that also higher order
corrections to the leading divergence are all $\propto 1/\sqrt{ak_{12}}$. Then we get
\beq
\mbox{log}~W_{\mbox{\scriptsize SYM}}\Big \vert_{spiky}~=~\frac{\Omega (\lambda )}{\sqrt{a~\vert \vec k_{1}-\vec k_2\vert }}~+~\Gamma_{\mbox{\scriptsize spike}}(\lambda)~\mbox{log}~a~+~{\cal O}(a^0)~.
\eeq

The function $\Omega (\lambda )$ would be
\beq 
\Omega (\lambda )~=~\lambda ~\frac{(\Gamma (1/4))^2}{2(2\pi )^{3/2}}~+~{\cal O}(\lambda ^2)\eeq
and
\beq
\Omega(\lambda)~=~\frac{8~\pi^{3/4}\sqrt{2\sqrt{2}+3}}{(\Gamma(\frac{1}{8}))^2}~\sqrt{\lambda}~+~\dots ~
\eeq
for small and large 't Hooft coupling, respectively. Use has been made of \eqref{weak-SYM},
\eqref{loop-malda} and {\it one half} of \eqref{spike-root}.

So far for $ \Gamma_{\mbox{\scriptsize spike}}(\lambda)$ we have only the weak coupling behavior \eqref{weak-SYM}
\beq
\Gamma_{\mbox{\scriptsize spike}}(\lambda)~=~\frac{\lambda}{2\pi^2}~+~{\cal O}(\lambda^2)~.
\eeq

Further study should confirm this picture by an analysis of next leading corrections.
Then the most interesting question is, whether $\Omega (\lambda ) $  and $\Gamma_{\mbox{\scriptsize spike}}(\lambda)$  are new independent functions of the coupling or whether they are somehow related to known ones. 

For QCD we have for both functions the weak coupling results. $\Omega (\lambda)$ is then half of its partner for SYM and $\Gamma_{\mbox{\scriptsize spike}}(\lambda)$ is zero.

It would be interesting to study also cases with cusp angle $\vartheta =\pi,~~\vec k_1=\vec k_2$ but a discontinuity in the third or higher derivative. Another straightforward generalisation concerns contours with additional discontinuities in the coupling to scalars on $S^5$.

Ignoring all divergences other than logarithmic ones, or equivalently using dimensional regularisation,  the function
$\Omega(\lambda)$ plays no role in the definition of the renormalised Wilson loop and the analysis of its scaling behaviour via the renormalisation group.
Then only $\Gamma_{\mbox{\scriptsize spike}}(\lambda)$ is of immediate interest.  But in any case, for the application to   entanglement entropy both functions
$\Omega(\lambda)$ and $\Gamma_{\mbox{\scriptsize spike}}(\lambda)$ are of physical interest.

A last comment concerns a comparison with another case, in which the coefficient in front of a logarithmic divergence becomes singular. The holographic
entanglement entropy  for static regions in $(3+1)$-dimensional CFT's has a logarithmic divergence. Its coefficient \cite{Solodukhin:2008dh,Dorn:2016bkd}
becomes singular, if the boundary
of the spatial region develops conical singularities. The direct calculation for such singular boundaries  yields a $\mbox{log}^2\epsilon$ divergences  \cite{Myers:2012vs, Klebanov:2012yf}. This has to be contrasted with our case,  where the transition to a stronger divergence yields $1/\sqrt{\epsilon}$.\\[20mm]
{\bf Acknowledgement:}\\[2mm]
I would like to thank Nadav Drukker for a helpful discussion and Florian Loebbert for a hint concerning the figures. 
\newpage
\section*{Appendix A}
Here we discuss the divergences for $a\rightarrow 0$ of the integral in \eqref{spsc},
denoting it in this appendix by $I$
\beq
I(a,l,\vec k_1,\vec k_2)~=~\int_0^l\frac{dr}{r}\int _0^{\pi/2}~d\varphi~\frac{1}{1-\sin 2\varphi +\frac{r^2}{4}(\vec k_1~\cos ^2\varphi -\vec k_2~\sin^2\varphi)^2 +a^2/r^2}~.\nonumber
\eeq
If $a\rightarrow 0,~ r\rightarrow 0$, the $\varphi$-integration becomes divergent at $\varphi =\pi/4$. Therefore, we put $\varphi = \pi/4 +\psi$ and can write $I$ as
\beq
I(a,l,\vec k_1,\vec k_2)~=~\int_0^l\frac{dr}{r}\int _{-\pi/4}^{\pi/4}~d\psi~\frac{1}{\psi^2g_0(\psi)+r^2g_2
(\psi)+a^2/r^2}~,
\label{I2}
\eeq
with 
\bea
g_0(\psi)&=&\frac{1-\cos 2\psi}{\psi^2}~=~2~+~{\cal O}(\psi ^2)~,\label{g0}\\
g_2(\psi)&=&\frac{1}{16}~\big ( (1-\sin 2\psi )^2\vec k_1^2+(1+\sin 2\psi)^2\vec k_2^2-2~(\cos 2\psi)^2 ~\vec k_1\vec k_2\big )\nonumber\\
&=&\frac{1}{16}(\vec k_1-\vec k_2)^2~+~ ~{\cal O}(\psi )~.\label{g2}                          
\eea
To get the leading divergence one can replace $g_0(\psi)$ and $g_2(\psi)$ by their values
at $\psi =0$. But more care is necessary for the next-leading term, hence we continue with
\bea
I&=&I_1~+~I_2~,\label{I1I2}\\
I_1&=&\int_0^l\frac{dr}{r}\int _{-\pi/4}^{\pi/4}~d\psi~\frac{1}{\psi^2g_0(0)+r^2g_2
(0)+a^2/r^2}~,\\
I_2&=&\int_0^l\frac{dr}{r}\int _{-\pi/4}^{\pi/4}~d\psi~\Big (\frac{1}{\psi^2g_0(\psi)+r^2g_2
(\psi)+a^2/r^2}~-~\frac{1}{\psi^2g_0(0)+r^2g_2
(0)+a^2/r^2}\Big )~.\nonumber
\eea

Let us start with $I_2$. After $a/r=y$ we get
\beq
I_2~=~\int _{a/l}^{\infty} \frac{dy}{y}\int _{-\pi/4}^{\pi/4}~d\psi~\frac{\psi ^2(g_0(0)-g_0(\psi))+\frac{a^2}{y^2}(g_2(0)-g_2(\psi))}{(\psi^2g_0(\psi)+\frac{a^2}{y^2}g_2(\psi)+y^2)(\psi^2g_0(0)+\frac{a^2}{y^2}g_2
(0)+y^2)}~.\label{I-2}
\eeq
In the $\psi$-integral the limit $a\rightarrow 0$ can be applied to the integrand, thereby eliminating $g_2$ and
leading with \eqref{g0} to
\beq
\int _{-\pi/4}^{\pi/4}~d\psi~\frac{2\psi ^2-1+\cos 2\psi}{(1-\cos2\psi+y^2)(2\psi ^2+y^2)}~.
\eeq
For $y=0$ this  integral is equal to $(\frac{4}{\pi}-1)$.  
Using this in \eqref{I-2} we get
\beq
I_2~=~\big (1-\frac{4}{\pi}\big )~\log~ a~+~{\cal O}(a^0 )~.\label{I2final}
\eeq

Now  we proceed with $I_1$. After performing the $\psi$-integration we get with \eqref{g0},\eqref{g2} and using the shorthand
\beq
k_{12}~=~\vert \vec k_1-\vec k_2\vert ~,\label{k12}
\eeq
as well as the substitution $r=\sqrt{\frac{a}{k_{12}}}~\frac{2}{x}$ and the abbreviation
\beq
b~=~\frac{\sqrt{2ak_{12}}}{\pi}\label{ba}
\eeq
\beq
I_1~=~\frac{4}{\pi b}\int_{\frac{\sqrt{2}\pi b}{lk_{12}}}^{\infty}\mbox{arctan}\Big (\frac{x}{b\sqrt{1+x^4}}\Big )~\frac{dx}{\sqrt{1+x^4}}~.
\label{I-1b}\eeq
Obviously the leading divergent term for $b\rightarrow 0$ is given by
\beq
I_1^{\mbox{\scriptsize leading}}~=~\frac{4}{\pi b}~\frac{\pi}{2}~\int _0^{\infty}\frac{dx}{\sqrt{1+x^4}}~.\label{Ilead}
\eeq
To catch also the next-leading divergent term, we have to be more careful. With
\beq
I_1~=~\frac{4}{\pi b} \hat I_1(b)\label{Ihat}
\eeq 
we find from \eqref{I-1b}
\beq
\frac{d \hat I_1(b)}{db}~=~-\frac{\sqrt{2}\pi}{lk_{12}}~\mbox{arctan}\Big (\frac{\sqrt{2}\pi}{lk_{12}}\Big )~\big (1+{\cal O}(b^4)\big )~-~\int_{\frac{\sqrt{2}\pi b}{lk_{12}}}^{\infty}\frac{xdx}{x^2+b^2(1+x^4)}~.
\eeq
The integral in the second term can be done explicitly, and we find altogether
\beq
\frac{d \hat I_1(b)}{db}~=~2~\log ~b~+~{\cal O}(b^0)~,
\eeq
and after integration
\beq
\hat I_1(b)~=~\hat I_1(0)~+~2b~\log~ b~+~{\cal O}(b)~.
\eeq
Using this in \eqref{Ihat} we get with \eqref{ba} after evaluation of the integral in \eqref{Ilead}
\beq
I_1~=~\frac{\sqrt{\pi}~(\Gamma (1/4))^2}{2~\sqrt{2ak_{12}}}~+~\frac{4}{\pi}~\log~ a~+~{\cal O}(a^0)~.
\eeq
Combining this with \eqref{I1I2}, \eqref{I2final} the final result is
\beq
I(a,l,\vec k_1,\vec k_2)~=~\frac{\sqrt{\pi}~(\Gamma (1/4))^2}{2~\sqrt{2a~\vert \vec k_{1}-\vec k_2\vert }}~+~\log~ a~+~{\cal O}(a^0)~.
\label{Ifinal}
\eeq
\section*{Appendix B}
In the holographic calculation we are interested in  the $\epsilon\rightarrow 0$ behaviour of \eqref{Aeps}. Inserting in the nominator under the square root a zero in
the form of the l.h.s. of \eqref{region} we get
\beq
A_{\epsilon}^{\pm}~=~\frac{2~r_0^2}{\sqrt{\epsilon}}~\int_{r_{\epsilon}^{\pm}}^{r_0}\frac{\sqrt{r-r_{\epsilon}^{\pm}}\sqrt{1-\epsilon~ f_{\pm}(r,,r_{\epsilon}^{\pm})}}{r^2~\sqrt{r_0^4-r^4}}~dr~,\label{A-app-B}
\eeq
with
\beq
f_{\pm}(r,r_{\epsilon}^{\pm})~=~ \frac{(M\pm \sigma(r))^2-(M\pm \sigma (r_{\epsilon}^{\pm}))^2}{r-r_{\epsilon}^{\pm}}~+~ (r+r_{\epsilon}^{\pm}) ~.\label{f}
\eeq
The lower boundaries $r_{\epsilon}^{\pm}$, defined by \eqref{region},
behave as
\beq
r_{\epsilon}^{\pm}~=~\epsilon~(M\pm L/2)^2~+~\epsilon^3~(M\pm L/2)^4~+~{\cal O}(\epsilon ^4)~.\label{repsas}
\eeq
In this estimate use has been made of \eqref{sigma-r} and \eqref{r0L}. 

Anticipating the source of the leading divergence, we split $A_{\epsilon}^{\pm}$ in two pieces
\beq
A_{\epsilon}^{\pm}~=~A_{\epsilon,\mbox{\scriptsize lead}}^{\pm}~+~A_{\epsilon,\mbox{\scriptsize rem}}^{\pm}~,\label{Asplit}
\eeq
with
\beq
A_{\epsilon,\mbox{\scriptsize lead}}^{\pm}~=\frac{2}{\sqrt{\epsilon~r_0}}~\int_{r_\epsilon^{\pm}/r_0}^{1}\frac{\sqrt{x-r_{\epsilon}^{\pm}/r_0}}{x^2\sqrt{1-x^4}}~dx\label{lead}
\eeq
and
\beq
A_{\epsilon,\mbox{\scriptsize rem}}^{\pm}~=~\frac{2}{\sqrt{\epsilon~r_0}}~\int_{r_\epsilon^{\pm}/r_0}^{1}\frac{\sqrt{x-r_{\epsilon}^{\pm}/r_0}}{x^2\sqrt{1-x^4}}~\Big (\sqrt{1-\epsilon~f_{\pm}(xr_0,r_{\epsilon}^{\pm})}~-~1\Big )~dx~.\label{rem}
\eeq
The explicit calculation of the integral in \eqref{lead} yields $\frac{\pi}{2}\sqrt{r_0/r_{\epsilon}^{\pm}}$ plus a third order polynomial in $r_{\epsilon}^{\pm}/r_0$, with coefficients given by generalised hypergeometric functions $_5F_4$ of the argument $(r_{\epsilon}^{\pm}/r_0)^4$.
Its expansion for small $\epsilon$ then results via \eqref{repsas} in
\beq
A_{\epsilon,\mbox{\scriptsize lead}}^{\pm}~=~\frac{\pi}{\epsilon~(M\pm L/2)}~-~\frac{4\sqrt{\pi}~\Gamma(\frac{7}{8})}{\Gamma(\frac{3}{8})}~\frac{1}{\sqrt{\epsilon~r_0}}~+~{\cal O}(\epsilon^{1/2})~.\label{Alead}
\eeq

For the estimate of $A_{\epsilon,\mbox{\scriptsize rem}}^{\pm}$ it is crucial that
$f_{\pm}(r,r_{\epsilon}^{\pm})$ in \eqref{f} and \eqref{rem} remains bounded for $\epsilon\rightarrow 0$, {\it unifomly}  with respect to $r$. This implies
\beq
A_{\epsilon,\mbox{\scriptsize rem}}^{\pm}~=~{\cal O}(\epsilon)~\cdot~A_{\epsilon,\mbox{\scriptsize lead}}^{\pm}\label{Arem}~.
\eeq
Finally, we get with \eqref{Asplit},\eqref{Alead},\eqref{Arem}, as well as the $r_0\leftrightarrow L$ relation \eqref{r0L} and some $\Gamma$-function arithmetics
\beq
A^{\pm}_{\epsilon}~=~\frac{\pi}{\epsilon (M\pm L/2)}~-~\frac{8\sqrt{2}~\pi^{7/4}\sqrt{2\sqrt{2}+3}}{(\Gamma(\frac{1}{8}))^2}~\frac{1}{\sqrt{\epsilon L}}~+~{\cal O}(\epsilon ^0)~.\label{Afinal}
\eeq
\section*{Appendix C}
The Wilson loop contour in section 3 is of self-touching type, the tips of two
spikes coincide. To compare it with the result for an isolated spike in section 2 
requires some comments. 

As for self-crossing contours one has to expect mixing
under renormalisation. In our case it would be mixing between $W$ and $\tilde W$,
where $\tilde W$ is the correlator of the Wilson loops for the two single
circles.  The contour near our double spike with the arrows pointing in the direction of the colour flux and increasing contour parameter is shown in figure 4.\\[5mm] 
\begin{center}
\begin{tikzpicture}[scale=0.4]
\draw [-<,very thick] (145mm,0mm) arc (270:160:50mm);
\draw [-<,very thick] (150mm,0mm) arc (270:340:100mm);
\draw [->,very thick] (150mm,0mm) arc (270:380:50mm);
\draw [->,very thick] (145mm,0mm) arc (270:200:100mm);
\draw (188mm,30mm) node {{\bf 2}};
\draw (232mm,30mm) node {{\bf 1}};
\draw (105mm,30mm) node {{\bf 3}};
\draw (65mm,30mm) node {{\bf 4}};
\end{tikzpicture}\\[10mm]
\end{center}
Figure 4: {\it Contour from section 3. To draw the flow
of colour and contour parameter\\
$~~~~~~~~~~~$ in an eye-catching manner, the common tip
of the spikes has been splitted. }\\[2mm]
The four
legs are numerated as in the figure, and below the double index indicates between which legs the
propagators run. As argued in section 2, cases where the propagator has both ends
on the same leg have to be included only for those legs on which the arrow points
away from the spiky point. Then we get
\beq
\mbox{log}~W~=~\frac{g^2C_F}{4\pi^2}~\Big (\sum_{i,j}\big ( I^{ij}_{\mbox{\scriptsize sc}}~-~I^{ij}_{\mbox{\scriptsize gl}}\big )~+~\dots\Big )~+~{\cal O}(g^4)~,
\eeq
with

\bea
 I^{12}_{\mbox{\scriptsize sc}}&=& I^{34}_{\mbox{\scriptsize sc}}~=~\frac{B}{\sqrt{2ak_{12}}}~+~\mbox{log}~a~,\nonumber\\
I^{22}_{\mbox{\scriptsize sc}}&=&I^{44}_{\mbox{\scriptsize sc}}~=~\mbox{log}~a~,\nonumber\\[2mm]
I^{13}_{\mbox{\scriptsize sc}}&=&I^{14}_{\mbox{\scriptsize sc}}~=~I^{23}_{\mbox{\scriptsize sc}}~=~I^{24}_{\mbox{\scriptsize sc}}~=~-\mbox{log}~a~,
\eea 
and
\bea
 I^{12}_{\mbox{\scriptsize gl}}&=& I^{34}_{\mbox{\scriptsize gl}}~=~-\frac{B}{\sqrt{2ak_{12}}}~-~\mbox{log}~a~,\nonumber\\
I^{22}_{\mbox{\scriptsize gl}}&=&I^{44}_{\mbox{\scriptsize gl}}~=~I^{13}_{\mbox{\scriptsize gl}}~=~I^{24}_{\mbox{\scriptsize gl}}~=~\mbox{log}~a~,\nonumber\\[2mm]
I^{14}_{\mbox{\scriptsize gl}}&=&I^{23}_{\mbox{\scriptsize gl}}~=~-\mbox{log}~a~.
\eea 
Remarkably now, in contrast to the single spike case, the log$a$ terms cancel both for the gluonic as well as the scalar
terms. Then we get {\it both} for QCD and SYM a pure $1/\sqrt{a}$ divergence to be attributed to the touching spikes.

\end{document}